\documentclass[12pt]{article}
\usepackage{color}
\usepackage{amsfonts,amssymb,amsmath}
\usepackage{theorem,cite}
\theorembodyfont{\rmfamily}

\numberwithin{equation}{section}

\definecolor{brique}{rgb}{.9,.2,0}
\definecolor{blvert}{rgb}{0,.8,.85}
\definecolor{vertcl}{rgb}{0,1,.7}

\newcommand{\bea}{\begin{eqnarray}}
\newcommand{\eea}{\end{eqnarray}}
\newcommand{\beano}{\begin{eqnarray*}}
\newcommand{\eeano}{\end{eqnarray*}}
\newcommand{\beq}{\begin{equation}}
\newcommand{\eeq}{\end{equation}}
\newcommand{\nonu}{\nonumber \\}

\newcommand{\hs}[1]{\hspace{#1 mm}}

    \def\cH{{\cal H}}    
        
        \def\cO{{\cal O}}

\def\fn{{\mathfrak n}}

\newcommand{\CC}{{\mathbb C}}

\newcommand{\II}{{\mathbb I}}

\newcommand{\TT}{{\mathbb T}}

\newcommand{\wh}[1]{\widehat{#1}}
\newcommand{\wt}[1]{\widetilde{#1}}
\newcommand{\mb}[1]{\hs{4}\mbox{#1}\hs{4}}

\newcommand{\half}{\frac{1}{2}}

\newcommand{\atopn}[2]{\genfrac{}{}{0pt}{}{#1}{#2}}
\newcommand{\up}{\uparrow}
\newcommand{\down}{\downarrow}

\DeclareMathOperator{\res}{Res}
\begin{document}
\renewcommand{\thefootnote}{\fnsymbol{footnote}}
\newpage
\pagestyle{empty}
\setcounter{page}{0}




\null

\vfill

\begin{center}

 {\LARGE  {\sffamily Generalized coordinate Bethe ansatz\\[1.2ex]
 for non diagonal boundaries}}\\[1cm]

\vspace{10mm}
  
{\Large N. Crampe$^{ab,}$\footnote{nicolas.crampe@univ-montp2.fr} 
and E. Ragoucy$^{c,}$\footnote{ragoucy@lapp.in2p3.fr}}\\[.21cm] 
{\large $^a$ Universit\'e Montpellier 2, Laboratoire Charles Coulomb UMR 
5221,\\
F-34095 Montpellier, France \\
$^b$ CNRS, Laboratoire Charles Coulomb, UMR 5221, F-34095 Montpellier
}\\[.42cm]
{\large $^{c}$ Laboratoire de Physique Th{\'e}orique LAPTH\\[.242cm]
 CNRS and Universit{\'e} de Savoie.\\[.242cm]
   9 chemin de Bellevue, BP 110, F-74941  Annecy-le-Vieux Cedex, 
France. }
\end{center}
\vfill\vfill

\begin{abstract}
    We compute the spectrum and the eigenstates of the open XXX model 
    with non-diagonal (triangular) boundary matrices. Since the 
    boundary matrices are not diagonal, the usual coordinate 
    Bethe ansatz does not work anymore, and we use a generalization of 
    it to solve the problem.
\end{abstract}

\vfill
\rightline{\texttt{arXiv:1105.0338}}
\rightline{LAPTH-015/11}

\newpage
\pagestyle{plain}
\renewcommand{\thefootnote}{\arabic{footnote}}
\setcounter{footnote}{0}
\section{Introduction}
The XXX model with periodic boundary conditions \cite{heis} is one of 
the most studied model in the realm of integrable systems. It was 
solved by Bethe \cite{bethe}, using what is now called the 
coordinate Bethe ansatz, and since then, many papers appeared on the 
subject. As for most of integrable models, the situation drastically 
changes when one considers the open case, i.e. when the boundary 
conditions are not periodic anymore. From the pioneer works of 
Cherednik \cite{cherednik} and Sklyanin \cite{sklyanin}, it is known 
that integrability of the model is preserved when the boundary conditions 
are coded by two independent matrices obeying the so-called reflection 
equation. Classification of such matrices amounts to  
classify  integrable boundary conditions. However, although the models 
are known to be integrable, there is no general answer to get the 
Hamiltonian eigenstates for generic integrable boundaries (even when 
the periodic ones are known).

For the XXX model 
and its $su(N)$ generalization, such classification of boundaries
has been done in 
\cite{VG,MRS,selene}. However, as mentioned, 
the models are still not solved for generic boundary matrices. The situation 
where the two matrices are diagonal is well-understood: the spectrum 
can be computed using analytical Bethe ansatz \cite{MN,doikou2,byebye}, the 
eigenfunctions can be obtained from coordinate \cite{Gau, Gau2,ABBBQ} or 
algebraic
\cite{sklyanin,DVGR,gama,BR2} Bethe ansatz\footnote{Strictly speaking, references 
\cite{doikou2,DVGR} deal with the XXZ model. One has to 
perform a $q\to1$ limit to get the XXX model.}, and correlation functions are well 
studied, see e.g. \cite{IzKo,fctCorr,fctCorr2}. 
The case of simultaneously diagonalizable matrices is done 
in \cite{lepetit} (see also \cite{gama}). 

The spectrum for some non diagonal cases, where the boundary 
parameters (entering the boundary matrices) obey relations, could 
be deduced from studies on XXZ models, as dealt in e.g. 
\cite{Cao,nepo}. Let us also mention \cite{tq} where numerical methods are used to 
get the spectrum and \cite{BK} where a deformation of Onsager algebra 
is studied to compute the spectrum for general boundary matrices. 
The first attempt to 
compute the spectrum and 
eigenfunctions for XXX model with boundary matrices that are not 
simultaneously diagonalizable can be found in \cite{MRM}. There, the algebraic Bethe ansatz 
is used but one
 must restrict itself to one triangular boundary and one diagonal. Moreover, the $su(2)$ invariance of the $R$-matrix is needed,
so that the treatment is specific to the XXX model.
The aim of this article is to tackle the same question but using a
generalized coordinate Bethe ansatz (gCBA). 
Indeed, recently, in the case of XXZ model 
such problem has been successfully solved using such a method \cite{simon09,dam}.
It leads to a simpler presentation of the gCBA, that we present from the integrable models view point 
(while the presentation in \cite{simon09,dam} was more ASEP like). We hope that 
this presentation permits to emphasize the novelties of the gCBA without the 
technical difficulties encountered in the ASEP model.
 The gCBA allows us to construct the eigenstates and compute the 
spectrum for one diagonal and one non-diagonal boundary matrix. 
The solution is equivalent to the $q\to1$ limit of 
the XXZ model \cite{Cao,nepo,dam} and the one found by algebraic Bethe ansatz \cite{MRM}.
The constraints on the boundary matrices for XXZ model found previously \cite{Cao,nepo,dam} 
being now replaced by the 
triangular form of the left boundary matrix and the diagonal form of the right 
one. 

To be more specific, we will construct the eigenfunctions and 
compute the spectrum and Bethe equations for the following Hamiltonian:
\beq
H=B_{1}^++H_{bulk}+B_{L}^- \mb{where} 
H_{bulk}=\sum_{\ell=1}^{L-1} h_{\ell,\ell+1}
=\sum_{\ell=1}^{L-1} \Big(P_{\ell,\ell+1}-\II\Big)
\label{eq:Hxxx}
\eeq
with the following boundary matrices
\beq
B^+=\left(\begin{array}{cc} \alpha & \mu \\ 0 & \beta 
\end{array}\right)
\mb{and} B^-=\left(\begin{array}{cc} \gamma & 0 \\ 0 & \delta 
\end{array}\right)\,.
\eeq
Let us stress that, generically, both boundary matrices 
are not simultaneously diagonalizable (or even not diagonalizable at 
all), since they do not commute and that the total spin is not a good quantum number, 
since it does not commute with the Hamiltonian (because of the triangular boundary matrix). 
Of course, the method also applies when $B^+ \leftrightarrow B^-$
or when one considers a lower triangular matrix instead of a 
upper triangular one. Moreover, one can also conjugate both $B^+$ and 
$B^-$ by the same matrix: the spectrum is the same, and the 
eigenstates are constructed in an obvious way.

The plan of the paper is as follows. 
Notations are detailed in section \ref{sec:modXXX}. The construction 
of the eigenstates is done in section \ref{sec:gCBA}, and section 
\ref{sec:proof} is devoted to the proof of the main property of the 
article. We conclude in section \ref{sec:conclu}.

\section{XXX model with boundaries\label{sec:modXXX}}
The model we consider has Hamiltonian (\ref{eq:Hxxx}). It acts on a 
spin chain with $L$ sites, with a spin $\half$ (a $\CC^2$ space)
on each site of the 
chain. Hence, the Hilbert space is $\cH=\big(\CC^2\big)^{\otimes L}$, 
and $H\in\mbox{End}(\cH)$. 

We will use the auxiliary space notation, where indices indicate on 
which sites (of the chain) the operators act non trivially. For instance, for an 
operator acting on two sites, $\cO\in \mbox{End}(\CC^2\otimes\CC^2)$, 
$\cO_{34}$ will denote the operator $\cO$ acting on sites 3 and 4 of 
the chain, i.e. 
\beq
\cO_{34}=\II\otimes\II\otimes\cO\otimes(\II)^{\otimes 
(L-4)}\in\mbox{End}(\cH)\,,
\eeq
 where $\II$ denotes the $2\times2$ 
identity matrix.

 Hence, in (\ref{eq:Hxxx}), $B_{1}^+$ is the left boundary 
matrix $B^+$ acting on the first site 1:
\beq
B_{1}^+=B^+\otimes\underbrace{\II\otimes\ldots\otimes\II}_{L-1}\,,
\eeq
 $h_{\ell,\ell+1}$ are local 
Hamiltonians acting on sites $(\ell,\ell+1)$
\beq
h_{\ell,\ell+1}=\underbrace{\II\otimes\ldots\otimes\II}_{\ell-1}\otimes h\otimes
\underbrace{\II\otimes\ldots\otimes\II}_{L-\ell-1}\,,
\eeq
and $B_{L}^-$ is the 
right boundary matrix $B^-$ acting on the last site $L$. 

$P_{\ell,\ell+1}$ is the permutation operator between site 
$\ell$ and site $\ell+1$. As a matrix, the permutation $P$ takes the form
\beq
P=\left(\begin{array}{cccc} 1 & 0 & 0 & 0 \\ 0 & 0 & 1 & 0 \\
0 & 1 & 0 & 0 \\ 0 & 0 & 0 & 1 \end{array}\right)\,.
\eeq

The Hamiltonian (\ref{eq:Hxxx}) describes the interaction of spins 
(up or down) among themselves, and with two boundaries describes by 
the matrices $B^\pm$. We are looking for its eigenstates and its 
spectrum. 

Contrary to the periodic case, the open XXX Hamiltonian does not 
possess an $su(2)$ symmetry, so that the spin cannot be used as a 
quantum number. For non-diagonal boundary matrices, it does not 
commute with the $u(1)$ generator, so that the pseudo-excitation number is 
not a good quantum number either.
We will come back on this point in the next section.

\subsubsection*{Reference state}
The state 
\beq
|\up\ldots\up\rangle
\in\big(\CC^2\big)^{\otimes L}
\label{eq:def:vac}
\eeq 
is an eigenstate:
\beq
H\, |\up\ldots\up\rangle = (\alpha+\gamma)\, |\up\ldots\up\rangle\,.
\eeq
This state will be chosen as a reference state (the so-called 
pseudo-vacuum), and states with some spin down will be considered as 
pseudo-excitations above this reference state, see eq. (\ref{eq:def:state}).
Let us stress that the reference state is \underline{not}, in general, the ground 
state, and that spins down are \underline{not} physical excitations. 
It is just a convenient way to parametrize all the states in $\cH$.

On contrary, due to the left boundary, the state $|\down\ldots\down\rangle$ is 
\underline{not} an eigenstate.

\subsubsection*{Hermiticity}
Note also that with this choice of boundaries, the Hamiltonian $H$ 
is not Hermitian anymore for generic values of the boundary parameters. 
However, we will see that the energy is the same as the one computed 
for diagonal boundary matrices, so that the energy is
real when the parameters $\alpha$, $\beta$, $\gamma$ and $\delta$ are 
real. 

Remark that $H$ can be pseudo-Hermitian:
\beq
H^\dag = U\, H\,U^\dag \mb{with}
U=\sigma^{x}\otimes \ldots\otimes \sigma^x
\mb{when} \alpha^*=\beta\mb{;} \delta^*=\gamma\mb{;} 
\mu^*=\mu\,,
\eeq
where $*$ denotes complex conjugation.
 This case is still outside the 
range solved by usual coordinate Bethe ansatz (for $\mu\neq0$).

\subsubsection*{Integrability}
The model we consider is integrable. It can be built from the 
following transfer matrix
\bea
t(u) &=& \mbox{Tr}_{0}\Big(K^+_{0}(u)\,T_{0,<1\ldots L>}(u)\, 
K^-_{0}(u)\,\wh T_{0,<1\ldots L>}(u)\Big)\,,\\
T_{0,<1\ldots L>}(u) &=& R_{0L}(u)\,\ldots R_{01}(u) 
\mb{with} R_{0\ell}(u) = P_{0\ell}+u\,,\\
\wh T_{0,<1\ldots L>}(u) &=& R_{01}(u)\,\ldots R_{0L}(u) 
\,,\\
K^-(u) &=& \II+u\,\wt B^- \mb{with} 
\wt B^-=\left(\begin{array}{cc} \gamma-\delta & 0 \\ 0 & 
\delta-\gamma \end{array}\right)
\,,\\[1.2ex]
K^+(u) &=& \II+u\,\wt B^+\,,\mb{with} 
\wt B^+=\left(\begin{array}{cc} \alpha-\beta & \mu \\ 0 & 
\beta-\alpha \end{array}\right)
\eea
where the $K^-(u)$ matrix (resp. $K^+(u)$ one) obey the reflection 
equation (resp. dual reflection one).
Standard calculations \cite{sklyanin} show that one recovers the 
Hamiltonian
\beq
\half \left.\frac{dt(u)}{du}\right|_{u=0} 
=H+(L-1-\frac{\alpha+\beta+\gamma+\delta}{2})\,\II\,.
\eeq

\section{Generalized Coordinate Bethe Ansatz\label{sec:gCBA}}
The fact that the left boundary $B^+$ is not diagonal anymore implies 
that the usual coordinate Bethe ansatz fails. Indeed, since $B^{+}$ is 
triangular, it can flip a spin down to spin up, which is interpreted 
as the annihilation of a pseudo-excitation (or equivalently as its 
transmission outside the system). Hence, one cannot consider an 
eigenfunction with a given (fixed) number of pseudo-excitations. This is 
also another way to see that the spin is not a `good' quantum number, 
since the Hamiltonian changes it.
However, since there 
is only annihilation, one can consider eigenfunctions having a fixed 
\textit{maximum} number of pseudo-excitations. It leads us to the following ansatz:
\begin{equation}
\label{eq:ansatz}
 \Phi_n=\sum_{m=0}^n\ \sum_{x_{m+1}<\dots<x_n}\ \sum_{g\in G_m}\
A_g^{(n,m)}\ e^{i\boldsymbol{k}^{(m)}_g.\boldsymbol{x}^{(m)}}\ 
|x_{m+1},\dots,x_n\rangle\,,
\end{equation}
where $G_m$ is a full set of representatives of the coset 
$BC_n/BC_m$ ($G_0=BC_n$, by convention) and $BC_{m}$ is the $B_{m}$ Weyl group, generated by  
transpositions $\sigma_{j}$, $j=1,\ldots,m-1$ that exchange $k_j$ 
and $k_{j+1}$, and the reflection $R_{1}$ exchanging $k_{1}$ and $-k_{1}$. 
The vectors $|x_{m+1},\dots,x_n\rangle$ are given by 
\beq
|x_{m+1},\dots,x_n\rangle=
|\up\ldots\up\!\!\raisebox{-1.2ex}{$\atopn{\displaystyle\down}{x_{m+1}}$}\!\!
\up\ldots\up\!\!\raisebox{-1.2ex}{$\atopn{\displaystyle\down}{x_{m+2}}$}\!\!
\up\ldots\ldots\up
\raisebox{-1.2ex}{$\atopn{\displaystyle\down}{x_{n}}$}\up\ldots\up\rangle
\in\big(\CC^2\big)^{\otimes L}
\label{eq:def:state}
\eeq 
and we introduce the notation $\boldsymbol{k}^{(m)}_{g}$ for the 
following truncated vector
\begin{equation}
\boldsymbol{k}^{(m)}_{g}=(k_{g(m+1)},\dots,k_{g(n)})\;.
\end{equation}
For this definition to be consistent, the coefficients $A_{g}^{(n,m)}$ 
do not have to depend on the choice of the representative i.e.
\begin{equation}
 A_{gh}^{(n,m)}=A_{g}^{(n,m)}\mb{for any $h\in BC_m$.}
\end{equation}

The coefficients $A_{g}^{(n,m)}$ are complex numbers to be determined 
such that $\Phi_n$ is
an eigenfunction of $H$ i.e. such that the following equation holds
\begin{equation}\label{eq:sch}
 H \Phi_n = E \Phi_n\;.
\end{equation}
We project equation (\ref{eq:sch}) on the different independent 
vectors to get constraints on the coefficients 
$A^{(n,m)}_{g}$.

Since $H$ is a sum of operators acting on (at most) two neighboring sites 
only, one has to single out the cases where the $x$'s obey the 
following constraints:
\begin{itemize}
\item  all the $x_{j}$'s are far away one from each other 
($1+x_j<x_{j+1}$, $\forall\,j$) and are not on the 
boundary sites 1 and $L$. This case will be called 
generic\footnote{Here and below, unless explicitly specified, all 
the (sub)sets of $x_{i}$'s will be considered as generic.}.
\item $x_{j}+1=x_{j+1}$ for some $j$,
\item $x_{1}=1$, or $x_{m}=L$.
\end{itemize}
As the eigenvalue problem is a linear problem, it is 
enough to treat the cases where at most one of the particular cases 
appears: more complicated cases just appear as superposition of  
`simple' ones. 

\subsubsection*{Calculation of the energy: projection on 
$\boldsymbol{|x_{1},\dots,x_n\rangle}$ for 
$\boldsymbol{(x_{1},\dots,x_n)}$ generic}\ 

As in the usual coordinate Bethe ansatz \cite{bethe}, this projection 
provides the energy:
\begin{equation} 
E=\alpha+\gamma+\sum_{j=1}^n\lambda(e^{ik_j})
\mb{where}
\lambda(u)=u+\frac{1}{u}-2=\frac{(u-1)^2}{u}\,.
\label{def:lambdau}
\end{equation}
Let us remark that, up to the boundary terms $\alpha$ and 
$\gamma$, the energy takes 
the same form  as in the periodic case. 

We want also to stress that the coefficient 
$\mu$ does not enter the energy, hence the spectrum for the model 
is the same as the spectrum of the model based on diagonal 
matrices\footnote{To be precise, one should also check that the Bethe 
equations do not depend on $\mu$ either: one will see below that it is indeed 
the case.\label{fo:nm}}.

\subsubsection*{Scattering matrix: projection on 
$\boldsymbol{|x_{1},\dots,x_{j},x_{j+1}=1+x_{j},x_{j+2},\dots,x_n\rangle}$}\ 

As usual, this projection provides the scattering 
matrix between pseudo-excitations. It is given by a relation between 
$A^{(n,0)}_{g}$ and $A^{(n,0)}_{g\sigma_j}$ where $\sigma_j$ is the 
permutation of $j$ and $j+1$.
Namely, we get
\begin{eqnarray} 
\label{eq:S}
&&A^{(n,0)}_{g\sigma_j}
=S\!\left(e^{ik_{gj}},e^{ik_{g(j+1)}}\right)\,A^{(n,0)}_{g}\,,
\\
&&\mb{with}
 S(u,v)=-\frac{a(u,v)}{a(v,u)}
\mb{and}
a(u,v)=i\,\frac{2v-uv-1}{uv-1}\,.
\label{def:alphau}
\end{eqnarray}
The normalization chosen for the function $a(u,v)$ is for 
further simplifications.
As expected, this relation is similar to the periodic case since the 
boundaries 
are not involved in this process.

\subsubsection*{First relation: projection on 
$\boldsymbol{|1,x_{m+1}\dots,x_n\rangle}$, $\boldsymbol{m\geq1}$}\

This relation is a new one. We get, for any $g\in G_m$,
\begin{eqnarray}\label{eq:1g}
\sum_{h\in H_m} A_{gh}^{(n,m-1)}e^{ik_{ghm}}
\Big(\beta-\alpha-1+e^{ik_{ghm}}
-\sum_{j=1}^{m}\lambda(e^{ik_{gj}})\Big)
=0\,,
\end{eqnarray}
where $H_m=BC_m/BC_{m-1}$.
To obtain this relation, we have used the following property (valid 
for any function $f$):
\begin{equation}
 \sum_{g\in G_{m-1}}f(g)
e^{ik_{g(m)}}e^{i\boldsymbol{k}_{g}^{(m)}\boldsymbol{x}^{(m)}}
=\sum_{g\in G_{m}}e^{i\boldsymbol{k}_{g}^{(m)}\boldsymbol{x}^{(m)}}
\sum_{h\in H_m}f({gh})\,e^{ik_{gh(m)}}\;.
\end{equation}
Let us stress that equation (\ref{eq:1g}) does not depend on the choice 
of representative of $G_m$.

\subsubsection*{Reflection coefficient for the left boundary}\ 

The reflection coefficient is computed using relation (\ref{eq:1g}) 
at $m=1$. In that case, $H_{1}=BC_{1}$ is constituted of 2 elements 
only: the identity $id$ and the reflection $R_{1}$ that changes $k_1$ into 
$-k_1$. One gets
\beq
A^{(n,0)}_{gR_{1}} = R(e^{ik_{g1}})\ A^{(n,0)}_{g}
\mb{with} 
R(u)=-u^2\,\frac{1-\frac1u+\beta-\alpha}{1-u+\beta-\alpha}
=\frac{r_+(1/u)}{r_+(u)}\,,
\label{eq:reflu}
\eeq
where $r_+(u)$ is given in (\ref{def:ru}). 

Remark that 
$R(u)\,R(\frac1u)=1$: for the \textit{physical} excitations, the left 
boundary is purely reflective (no loss of excitations).

\subsubsection*{Second relation: projection on 
$\boldsymbol{|x_{m+1}\dots,x_n\rangle}$, $\boldsymbol{m\geq 1}$}\ 

This projection provides a relation between the coefficients 
from the level $m-1$ and $m$ 
since we must take into account that the left boundary can destroy one
pseudo-excitation present on the site 1.
We obtain the following constraint, for any $g\in G_m$,
\begin{equation}\label{eq:2g}
 \mu\,\sum_{h\in H_m}A_{gh}^{(n,m-1)}e^{ik_{gh(m)}}
-\sum_{j=1}^{m}A_g^{(n,m)}\,\lambda(e^{ik_{gj}})=0
\,.
\end{equation}

\subsubsection*{Transmission coefficient for the left boundary}\ 

{From} (\ref{eq:1g}) and (\ref{eq:2g}), we may express all the 
$A_g^{(n,m)}$ ($m\geq 1$) in terms of $A_g^{(n,0)}$
thanks to the recursive relation
\begin{equation}
\label{eq:recuT}
 A_g^{(n,m)}=T^{(m)}(e^{ik_{g1}},\dots,e^{ik_{gm}})\,A_g^{(n,m-1)}
\mb{for} 1\leq m\leq n\,,
\end{equation}
with the following definitions:
\bea
T^{(m)}(u_{1},\dots,u_{m}) =\frac{\mu}{r_+(u_{m})}\
\left( \prod_{j=1}^{m-1}a(u_{m},u_j)\,a(u_{j},1/u_m)\right)^{-1} 
\,,\label{eq:Tmu}\\
r_+(u) =\lambda(u)\,\frac{1-u+\beta-\alpha}{1-u^2}=
-\frac{(u-1)(1-u+\beta-\alpha)}{u(1+u)}\,.
\label{def:ru} 
\eea
Relation (\ref{eq:recuT}) can be interpreted as the transmission of 
one (among $m$) pseudo-excitation(s) through the left boundary: this 
pseudo-excitation has been destroyed (from the spin chain view point). 

Remark that $T^{(m)}$ is proportional to $\mu$, in accordance with the 
picture described at the beginning of the section: when $\mu=0$, the 
boundary becomes diagonal, the usual coordinate Bethe ansatz works, 
and there is no relation between $A^{(n,m)}_{g}$ and $A^{(n,m-1)}_{g}$, 
each level $m$ providing an independent eigenfunction (with a fixed number 
of pseudo-excitations); when $\mu\neq0$ all the levels $m$ are related, but 
we get an independent eigenfunction for each upper number $n$ of 
pseudo-excitations.\\

Iterative use of (\ref{eq:recuT}) leads to
\bea
A_g^{(n,m)} &=& \TT^{(m)}(e^{ik_{g1}},\dots,e^{ik_{gm}})\,A_g^{(n,0)}\,,\\
\TT^{(m)}(u_{1},\dots,u_{m}) &=& 
\frac{\mu^m}{r_+(u_{m})\,r_+(u_{m-1})\ldots r_+(u_{1})}\
\left( \prod_{1\leq j<\ell\leq m} a(u_{\ell},u_j)\,a(u_{j},1/u_\ell) 
\right)^{-1}
\label{eq:Ttot}
\eea

The proof that (\ref{eq:recuT}-\ref{eq:Tmu}) is a solution of 
both equations (\ref{eq:1g}) and (\ref{eq:2g}) is postponed to  
section \ref{sec:proof} and relies on a residue computation.
The integrability of the model plays a role at this place, since there 
are a priori too many constraints but not all of them are 
independent.\\

Note that relations (\ref{eq:S}), (\ref{eq:reflu}) and (\ref{eq:Ttot}) 
show that all the coefficients $A_g^{(n,m)}$ can be expressed in term 
of $A_{id}^{(n,0)}$. It is important to notice that this computation is consistent since the
obtained expression does not depend 
on the way we write $g$ in terms of the generators $\sigma_j$ and $R_1$.

\subsubsection*{Bethe equations: projection on 
$\boldsymbol{|x_{1}\dots,x_{n-1},L\rangle}$} 

This last constraint consists in the quantization of the pseudo-excitations moments 
since the 
system is in a finite volume. In the context of the coordinate Bethe 
ansatz,
this quantization leads to the so-called Bethe equations, explicitly 
given by
\begin{eqnarray}
&& \prod_{\substack{\ell=1 \\ \ell\neq j}}^n 
S(e^{ik_\ell},e^{ik_j})\,S(e^{-ik_j},e^{ik_\ell})
=e^{2iLk_j}\,
\frac{r_+(e^{ik_{j}})\,r_-(e^{ik_{j}})}
{r_+(e^{-ik_{j}})\,r_-(e^{-ik_{j}})} \mb{for} 1\leq j \leq n
\label{eq:bethe}\\
&& r_{-}(u) = \frac{u-1}{u+1}\,(1-u+\delta-\gamma)\,.
\label{def:rLu}
\end{eqnarray}
We remind that the scattering matrix $S(u,v)$ is given in (\ref{def:alphau}), 
while the $r_{+}(u)$ function is defined in (\ref{def:ru}).

Let us remark that Bethe 
equations (\ref{eq:bethe}) do not depend on the parameter $\mu$. Therefore,
as previously mentioned, the 
spectrum (i.e. the energy (\ref{def:lambdau})) is also independent of this parameter.
Therefore, the spectrum is similar to the one with diagonal boundaries ($\mu=0$). 
In \cite{MRM}, similar results on the spectrum have been obtained via algebraic Bethe ansatz.
In contrary, the eigenvectors depends on the parameter $\mu$ via relation (\ref{eq:Ttot}) determining 
the coefficients entering in our ansatz. The eigenvectors computed thanks to the algebraic Bethe ansatz
in \cite{MRM} also depends  on $\mu$ since the creation operators $\wt B(\lambda)$, used to construct the ansatz, 
depends on $\mu$. Unfortunately, a direct proof that the eigenvectors constructed via both methods are identical
is a difficult task and beyond the scope of this paper. 
  
\subsubsection*{Completeness of the ansatz}\
It is clear that two states $\Phi_{n_{1}}$ and $\Phi_{n_{2}}$ are  
independent when $n_{1}\neq n_{2}$. Thus it remains to prove that the 
Bethe equations provide the right number of solutions, and that, at 
given $n$, these solutions are independent.

Moreover, it is believed that the coordinate Bethe ansatz for open XXX 
chain\footnote{To our knowledge it is only proven for closed spin 
chains, see e.g. \cite{Kiri,Baxter}.} 
is complete when $\mu=0$. Since the 
Bethe equations do not depend on $\mu$, they provide the same number of 
solutions. Hence, from the conjecture at $\mu=0$, we deduce that when 
$\mu\neq0$, the set of solutions has the right dimension to get the 
complete set of eigenstates. 

\pagebreak[4]

\section{Proof of the transmission relation (\ref{eq:recuT})\label{sec:proof}}
In this section, we prove that (\ref{eq:recuT}) implies (\ref{eq:1g}) and 
(\ref{eq:2g}).\\

We start with (\ref{eq:2g}). First, we remark that a consequence 
of 
(\ref{eq:recuT}) is
\begin{equation}
\label{eq:Sm}
 A_{g\sigma_j}^{(n,m)}=
A_{g}^{(n,m)}\times\begin{cases}
 1 &1\leq j \leq m-1\,,\\
\frac{T^{(m)}(e^{ik_{g1}},\dots,e^{ik_{g(m-1)}},e^{ik_{g(m+1)}})}
{T^{(m)}(e^{ik_{g1}},\dots,e^{ik_{gm}})}\
S(e^{ik_{gm}},e^{ik_{g(m+1)}})& j= m\,,\\
S(e^{ik_{gj}},e^{ik_{g(j+1)}})& j\geq m+1\,.
\end{cases}
\end{equation}
Then, using again (\ref{eq:recuT}) to express now
$A_g^{(n,m+1)}$ in terms of $A_g^{(n,m)}$
and using (\ref{eq:Sm}) to express $A_{gh}^{(n,m)}$ in terms of 
$A_g^{(n,m)}$, relation 
(\ref{eq:2g}) becomes
the functional relation
\begin{eqnarray}\label{eq:ss1}
\sum_{j=1}^{m+1}\left[u_j\,r_+(u_j)
\prod_{\substack{\ell=1\\ \ell\neq 
j}}^{m+1}a(u_j,u_\ell)\,a(u_\ell,\frac{1}{u_j})
+\frac{1}{u_j}\,r_+(\frac{1}{u_j})
\prod_{\substack{\ell=1\\ \ell\neq 
j}}^{m+1}a(\frac{1}{u_j},u_\ell)\,a(u_\ell,u_j)
-\lambda(u_j)\right]
 =0\,,
\end{eqnarray}
where $u_j$ stands for $\exp(ik_{gj})$ and the functions are defined 
in (\ref{def:alphau}) and (\ref{def:ru}).

To prove this last relation (\ref{eq:ss1}), let us introduce the 
following function
\begin{equation}
 F^{(m)}(u)=\frac{1-u+\beta-\alpha}{2(1-u)}\,
\prod_{\ell=1}^ {m} a(u,u_\ell)\,a(u_\ell,\frac{1}{u})\,.
\end{equation}
Looking at the poles of $F^{(m)}(u)$, one can compute the residues of 
this function:
\begin{eqnarray}
\res(F^{(m)}(u))\Big|_{u=u_j} &=& u_j\,r_+(u_j)
\prod_{\substack{\ell=1\\ \ell\neq 
j}}^{m+1}a(u_j,u_\ell)\,a(u_\ell,\frac{1}{u_j})
\,,
\\
\res(F^{(m)}(u))\Big|_{u=1/u_j} &=& 
\frac{1}{u_j}\,r_+(\frac{1}{u_j})
\prod_{\substack{\ell=1\\ \ell\neq 
j}}^{m+1}a(u_j,u_\ell)\,a(u_\ell,\frac{1}{u_j})
\,,
\\
\res(F^{(m)}(u))\Big|_{u=1}
&=& -\half(\beta-\alpha)\,,
\\
\res(F^{(m)}(u))\Big|_{u=\infty}
&=& \half(\beta-\alpha)
-\sum_{\ell=1}^{m+1}\lambda(u_\ell)
\qquad
\end{eqnarray}
Then, (\ref{eq:ss1}) is equivalent to $\sum_{residue} F^{(m)}(u)=0$ 
which proves (\ref{eq:2g}).

{From} (\ref{eq:1g}), we use the same procedure to obtain a new 
functional relation. After use of (\ref{eq:ss1}), this relation simplifies to
\begin{eqnarray}\label{eq:ss2}
&&\sum_{j=1}^{m+1}\left[u_j^2\,r_+(u_j)
\prod_{\substack{\ell=1\\ \ell\neq 
j}}^{m+1}a(u_j,u_\ell)\,a(u_\ell,\frac{1}{u_j})
+\frac{1}{u_j^2}\,r_+(\frac{1}{u_j})
\prod_{\substack{\ell=1\\ \ell\neq 
j}}^{m+1}a(\frac{1}{u_j},u_\ell)\,a(u_\ell,u_j)\right]
\nonu
&&+\Big( \beta-\alpha-1-\sum_{j=1}^m\lambda(u_j) \Big)
\Big(\sum_{k=1}^m\lambda(u_k) \Big)
 =0\,.
\end{eqnarray}
Again, this relation is equivalent to a residue calculation, the function 
to consider being now $G^{(m)}(x)=x\,F^{(m)}(x)$.

\section{Conclusion\label{sec:conclu}}
We have computed the spectrum and the eigenfunctions of the open XXX 
model with one triangular boundary matrix using a generalized coordinate 
Bethe ansatz.
The next step would be
 to compute scalar products of the eigenstates to get informations 
on the correlation functions of this  model. As far as scalar 
products are concerned, we think that an approach \textit{\`a la} Gaudin \cite{Gau} 
should work. A determinant formula, similar to the Slavnov 
determinant \cite{Slav} will be needed to compute the correlation 
functions in a simple way.

It remains to treat the case where the two boundary matrices are both 
triangular, and also the cases where one or two of the boundaries are 
general $2\times2$ matrices. Work is in progress on these two cases.
We expect to find constrains between the boundary parameters in the 
first case. We remind that the method has been applied successfully 
in the case of open XXZ model \cite{dam}.
The second case needs a further generalization of the 
coordinate Bethe ansatz.

Finally, we believe also that the method presented here can be useful to
find the spectrum for other integrable models with boundaries.



\begin{thebibliography}{99}
\bibitem{heis}
W.~Heisenberg,
\textsl{Zur Theorie des Ferromagnetismus,}
Zeitschrift f{\"u}r Physik {\bf 49} (1928) 619.

\bibitem{bethe}
H.~Bethe,
\textsl{Zur Theorie der Metalle. Eigenwerte und Eigenfunktionen 
Atomkete,}
Zeitschrift f{\"u}r Physik \textbf{71} (1931) 205.

\bibitem{cherednik}
I.V.~Cherednik, \textsl{Factorizing particles on a half line and root
systems,} Theor. Math. Phys. \textbf{61} (1984) 977.

\bibitem{sklyanin}
E.K. Sklyanin, \textsl{Boundary conditions for integrable quantum 
systems,}
J. Phys. \textbf{A21} (1988) 2375.

\bibitem{VG}
H.J. de Vega and A. Gonz{\'a}lez Ruiz,
\textsl{Boundary K-Matrices for the Six Vertex and the n(2n-1) $A_{n-1}$ Vertex Models,}
J.Phys. \textbf{A26} (1993) L519 and \texttt{arXiv:hep-th/9211114}.
 
\bibitem{MRS}
M.~Mintchev, E.~Ragoucy and P.~Sorba,
\textsl{Spontaneous symmetry breaking in the gl(N)-NLS hierarchy on 
the half line},
J.\ Phys. {\bf A34} (2001) 8345
and \texttt{hep-th/0104079}.
  
\bibitem{selene} D. Arnaudon, J. Avan, N. Crampe, A. Doikou, L. 
Frappat and E.
Ragoucy, \textsl{General boundary conditions for the $sl(N)$ and 
$sl(M|N)$ open spin chains}, JSTAT \textbf{08} (2004) P005 and 
\texttt{arXiv:math-ph/0406021}.

\bibitem{MN}
L. Mezincescu and R.I. Nepomechie,
\textsl{Analytical Bethe Ansatz for quantum-algebra-invariant spin chains,} 
Nucl. Phys. \textbf{B372} (1992) 597 and \texttt{hep-th/9110050}.

\bibitem{doikou2}
A.~Doikou, \textsl{Fusion and analytical Bethe Ansatz for the
$A_{n-1}^{(1)}$ open spin chain,} J. Phys. \textbf{A33} (2000) 4755 and
\texttt{arXiv:hep-th/0006081}.

\bibitem{byebye} D. Arnaudon, N. Crampe, A. Doikou, L. Frappat
and E. Ragoucy,
\textsl{Analytical Bethe ansatz for closed and open $gl(n)$-spin 
chains in
any representation,} JSTAT \textbf{02} (2005) P02007 and
\texttt{arXiv:math-ph/0411021}.

\bibitem{Gau}
M. Gaudin, \textsl{La fonction d'onde de Bethe},
Masson, Paris (1983).

\bibitem{Gau2} M. Gaudin,
\textsl{Boundary Energy of a Bose Gas in One Dimension,}
Phys. Rev. {\bf A4} (1971) 386.

\bibitem{ABBBQ} 
F.C.~Alcaraz, M.N.~Barber, M.T.~Batchelor, R.J.~Baxter,G.R.W.~Quispel,
\textsl{Surface exponents of the quantum XXZ, Ashkin-Teller and Potts model,}
J. Phys. \textbf{A20} (1987) 6397.

\bibitem{DVGR}
H.J.~de Vega and A.~Gonz{\'a}lez-Ruiz, \textsl{Exact solution of the
$SU_{q}(n)$ invariant quantum spin chains,} Nucl. Phys. \textbf{B417}
(1994) 553 and \texttt{arXiv:hep-th/9309022};\\
\textsl{Exact Bethe ansatz solution of 
$A_{\fn-1}$ chains with non- $SU_q(\fn)$ invariant open boundary conditions},
\texttt{arXiv:hep-th/9404141}.

\bibitem{gama} W. Galleas and M.J. Martins,
\textsl{Solution of $su(N)$ vertex model with non-diagonal open 
boundaries}, Phys. Lett. \textbf{A335} (2005) 167 and \texttt{nlin.SI/0407027}. 
  
\bibitem{BR2} 
S.~Belliard and {\'E}.~Ragoucy,
\textsl{Nested Bethe ansatz for `all' open spin chains with diagonal 
boundary conditions}, J. Phys. \textbf{A42} (2009) and
\texttt{arXiv:0902.0321}.

\bibitem{IzKo} A.G. Izergin and V.E. Korepin,
\textsl{The quantum inverse scattering approach to correlation functions,}
Comm. Math. Phys. \textbf{94} (1984) 67.

\bibitem{fctCorr}
O.A. Castro-Alvaredo and J.M. Maillet, \textsl{Form factors of integrable 
Heisenberg (higher) spin chains}, J. Phys. \textbf{A40} (2007) 
7451 and \texttt{arXiv:hep-th/0702186}.

\bibitem{fctCorr2}
T. Deguchi and C. Matsui, 
\textsl{Correlation functions of the integrable 
higher-spin XXX and XXZ chains through the fusion method}, Nucl. 
Phys. \textbf{B831} (2010) 359 and \texttt{arXiv:0907.0582}.



\bibitem{lepetit}
N.~Crampe, \textsl{Approches alg\'ebriques dans les syst\`emes
int\'egrables,} University of Savoie PhD Thesis (June 2004), in
French, Preprint LAPTH-These-1056/04, 
\texttt{http://lapth.in2p3.fr/lapth/preprint\_lapth/LAPTH1050.ps.gz}. 

\bibitem{Cao} 
J. Cao, H. Lin, K. Shi and Y. Wang, 
\textsl{Exact solutions and elementary excitations in the XXZ spin 
chain with unparallel boundary fields}, Nucl. Phys. \textbf{B663} (2003) 487 
and \texttt{arXiv:cond-mat/0212163}.
  
\bibitem{nepo} 
R.I. Nepomechie, 
\textsl{Bethe Ansatz solution of the open XXZ spin chain with 
nondiagonal boundary terms},
J.Phys. \textbf{A34} (2001) 9993 and \texttt{arXiv:hep-th/0110081};\\
R.I. Nepomechie,  \textsl{Functional relations and Bethe 
Ansatz for the XXZ chain},
J. Statist. Phys. \textbf{111} (2003) 1363 and 
\texttt{arXiv:hep-th/0211001};\\ 
L. Frappat, R. Nepomechie and E. Ragoucy,
\textsl{Complete Bethe Ansatz solution of the open spin-s XXZ chain 
with  general integrable boundary terms},
JSTAT \textbf{0709} (2007) P09009 and
\texttt{arXiv:0707.0653}.

\bibitem{tq} H. Frahm, J. Grelik, A. Seel and T. Wirth, 
\textsl{Functional Bethe ansatz methods for the open XXX chain},
J. Phys. \textbf{A44} (2011) 015001 and \texttt{arXiv:1009.1081}.

\bibitem{BK}
P.~Baseilhac and K.~Koizumi,
\textsl{Exact spectrum of the XXZ open spin chain from the q-Onsager 
algebra
representation theory}, JSTAT \textbf{0709} (2007) P09006 and
\texttt{arXiv:hep-th/0703106};
\\
P.~Baseilhac,
\textsl{New results in the XXZ open spin chain}, proceedings of 
RAQIS07 (Annecy-le-Vieux, France) and \texttt{arXiv:0712.0452}.

\bibitem{MRM} C.S.~Melo, G.A.P.~Ribeiro and M.J.~Martins,
\textsl{Bethe ansatz for the XXX-S chain with non-diagonal open boundaries,}
Nucl. Phys \textbf{711} (2005) 565 and \texttt{nlin/0411038}.


\bibitem{simon09}
D. Simon, \textsl{Construction of a coordinate Bethe Ansatz for the 
asymmetric exclusion process with open boundaries}, J. Stat. Mech. 
(2009) P07017 and \texttt{arXiv:0903.4968}.

\bibitem{dam}  
N. Crampe, E. Ragoucy and D. Simon,
\textsl{Eigenvectors of open XXZ and ASEP models for a class of 
non-diagonal boundary conditions}, JSTAT \textbf{1011} (2010) P11038 and
\texttt{arXiv:1009.4119}.

\bibitem{Kiri} A. N. Kirillov,
\textsl{Combinatorial identities, and completeness of eigenstates of 
the Heisenberg magnet,}
Zap. Nauch. Sem. LOMI \textbf{131} (1984) 88.

\bibitem{Baxter} R.J. Baxter, 
\textsl{Completeness of the Bethe ansatz for the six and eight-vertex 
models}, J. Stat. Phys. \textbf{108} (2002) 1-48
and \texttt{arXiv:cond-mat/0111188}.

\bibitem{Slav} N.A. Slavnov, 
\textsl{Calculation of scalar products of wave 
functions and form factors
in the framework of the algebraic Bethe ansatz}, 
Theor. Math. Phys. {\bf 79} (1989) 502-508.

\end{thebibliography}
\end{document}